\documentclass[prl,twocolumn,amsmath,amssymb,superscriptaddress]{revtex4-1}

\usepackage[T1]{fontenc} 
\usepackage{bm}
\usepackage{graphicx}
\usepackage[bookmarks=false]{hyperref}
\usepackage{color}

\newcommand{\bra}[1]{\langle #1|}
\newcommand{\ket}[1]{|#1\rangle}
\newcommand{\braket}[1]{\langle #1 \rangle}

\def\ee{\mathrm{e}}
\def\ii{\mathrm{i}}

\def\const{\mathrm{const.}}

\def\ketg{\ket{0}}
\def\brag{\bra{0}}

\def\dX{\Delta X}
\def\dXmin{\dX_{\text{min}}}
\def\dXmax{\dX_{\text{max}}}

\def\rabi{g}
\def\wbt{\tilde{\omega}_b}
\def\rabit{\tilde{g}}
\def\DDt{\tilde{D}}
\def\wbc{\check{\omega}_b}
\def\rabic{\check{g}}
\def\HH{\mathcal{H}}

\def\thetaopt{\theta_{\text{opt}}}
\def\psiopt{\psi_{\text{opt}}}
\def\phiopt{\varphi_{\text{opt}}}

\def\oHH{\hat{\mathcal{H}}}
\def\oHHDicke{\oHH_{\text{Dicke}}}
\def\oS{\hat{S}}
\def\oU{\hat{U}}
\def\oUd{\hat{U}^{\dagger}}
\def\oUz{\hat{U}_{0}}
\def\oUzd{\oUz^{\dagger}}

\def\oa{\hat{a}}
\def\oad{\hat{a}^{\dagger}}

\def\ob{\hat{b}}
\def\obd{\hat{b}^{\dagger}}
\def\oc{\hat{c}}
\def\ocd{\hat{c}^{\dagger}}
\def\od{\hat{d}}
\def\odd{\hat{d}^{\dagger}}

\def\op{\hat{p}}
\def\opd{\hat{p}^{\dagger}}
\def\oX{\hat{X}}

\def\ba{\bar{a}}
\def\bb{\bar{b}}

\begin{document}


\title{Perfect Intrinsic Squeezing at the Superradiant Phase Transition Critical Point}

\author{Kenji Hayashida}
\affiliation{Department of Electrical and Computer Engineering, Rice University, Houston 77005, USA}
\affiliation{Division of Applied Physics, Graduate School and Faculty of Engineering, Hokkaido University, Sapporo, Hokkaido 060-8628, Japan}
\author{Takuma Makihara}
\affiliation{Department of Physics and Astronomy, Rice University, Houston 77005, USA}
\author{Nicolas Marquez Peraca}
\affiliation{Department of Physics and Astronomy, Rice University, Houston 77005, USA}
\author{Diego Fallas Padilla}
\affiliation{Department of Physics and Astronomy, Rice University, Houston 77005, USA}
\author{Han Pu}
\affiliation{Department of Physics and Astronomy, Rice University, Houston 77005, USA}
\author{Junichiro Kono}
\affiliation{Department of Electrical and Computer Engineering, Rice University, Houston 77005, USA}
\affiliation{Department of Physics and Astronomy, Rice University, Houston 77005, USA}
\affiliation{Department of Material Science and NanoEngineering, Rice University, Houston 77005, USA}
\author{Motoaki Bamba}
\altaffiliation{E-mail: bamba.motoaki.y13@kyoto-u.jp}
\affiliation{Department of Physics, Kyoto University, Kyoto 606-8502, Japan}
\affiliation{PRESTO, Japan Science and Technology Agency, Kawaguchi 332-0012, Japan}
\date{\today}

\begin{abstract}
The ground state of the photon--matter coupled system described by the Dicke model is found to be perfectly squeezed at the quantum critical point of the superradiant phase transition (SRPT). In the presence of the counter-rotating photon--atom coupling, the ground state is analytically expressed as a two-mode squeezed vacuum in the basis of photons and atomic collective excitations.
The variance of a quantum fluctuation in the two-mode basis
vanishes at the SRPT critical point,
with its conjugate fluctuation diverging, ideally satisfying the Heisenberg uncertainty principle.
\end{abstract}


\maketitle
When photons strongly couple with an ensemble of atoms, there is a threshold coupling strength above which a static photonic field (i.e., a transverse electromagnetic field) and a static atomic field (i.e., an electric polarization) are expected to appear spontaneously. This phenomenon, known as the superradiant phase transition (SRPT)~\cite{Hepp1973AP,Wang1973PRA}, can occur not only at finite temperatures but also at zero temperature.  Since its first proposal in 1973, the SRPT has long attracted considerable attention from both experimental and theoretical researchers. In addition to the experimental demonstration of nonequilibrium SRPT in atoms confined in optical cavities \cite{Baumann2010N,barrett}, the possibilities of realizing the original SRPT in thermal equilibrium~\cite{Griesser2016PRA,Nataf2019a,Andolina2020} using various physical platforms including a superconducting circuit~\cite{Bamba2016circuitSRPT} and a magnonic system~\cite{Bamba2020}, have been demonstrated theoretically in recent years.

Although the finite-temperature SRPT is a classical phase transition in the sense that it is driven by thermal fluctuations \cite{Larson2017JPA,Note1},
quantum aspects of the SRPT have been investigated in terms of quantum chaos \cite{Emary2003PRL,Emary2003PRE},
entanglement entropy \cite{Lambert2004PRL},
and individual photonic and atomic squeezings \cite{Emary2003PRE,Castanos2011,Garbe2017PRA,Shapiro2019},
i.e., quantum fluctuation of the photonic (atomic) field is suppressed in one quadrature
whereas its conjugate fluctuation is enlarged
while satisfying the Heisenberg uncertainty principle.

Due to the ultrastrong photon--atom coupling \cite{Ciuti2005PRB,Kockum2018,Forn-Diaz2018},
which means that the coupling strength (vacuum Rabi splitting; anti-crossing frequency) is a considerable fraction of photonic and atomic resonance frequencies
and is required for realizing the SRPT,
it is known that the ground state of the photon--matter coupled systems
becomes a two-mode squeezed vacuum
\cite{Artoni1991,Artoni1989,Schwendimann1992,Schwendimann1992a,quattropani05,Ciuti2005PRB}
even in the normal phase (zero expectation values of the photonic and atomic fields).
One might expect that some critical quantum behavior
should exhibit at the onset of the SRPT.
However, how the squeezing property of the system would behave at the critical point is not well understood.


In the present letter, we show that {\em perfect} squeezing
is obtained in a photon--atom two-mode basis at the SRPT intrinsically,
i.e., in the ground state of the coupled system.
It means that, in contrast to the usual squeezing generation
in dynamical and nonequilibrium situations \cite{meystre99,walls08},
the SRPT can provide strong squeezing
stably in equilibrium situations
and has a potential to develop decoherence-robust quantum technologies.

We consider the isotropic Dicke model \cite{Dicke1954PR,Note2}
whose Hamiltonian is given by
\begin{equation} \label{eq:oHH_Dicke} 
\oHHDicke/\hbar
= \omega_a\oad\oa + \omega_b\left(\oS_z + \frac{N}{2}\right) + \frac{2\rabi}{\sqrt{N}}(\oad+\oa)\oS_x.
\end{equation}
Here, $\oa$ is the annihilation operator of a photon
with a resonance frequency $\omega_a$.
$\oS_{x,y,z}$ are the spin-$\frac{N}{2}$ operators
representing an ensemble of $N$ two-level atoms
with a transition frequency $\omega_b$.
$\rabi$ is the coupling strength and assumed to be real and positive 
for simplicity \cite{Note3}.
In terms of the lowering and raising operators
$\oS_{\pm} \equiv \oS_x \pm \ii \oS_y = \{\oS_{\mp}\}^{\dagger}$,
the last term (photon--atom coupling term) in Eq.~\eqref{eq:oHH_Dicke} can be rewritten as
${2\rabi}(\oad+\oa)\oS_x/{\sqrt{N}}
= {\rabi}(\oad+\oa)(\oS_+ + \oS_-)/{\sqrt{N}}.$
Among these four terms,
$\oad\oS_-$ and $\oS_+\oa$ are called the co-rotating terms
and responsible for the vacuum Rabi splitting; 
while
$\oad\oS_+$ and $\oa\oS_-$ are called the counter-rotating terms
and responsible for the vacuum Bloch-Siegert shift \cite{Forn-Diaz2010PRL,Li2018}. As we will show later, it is these counter-rotating terms that are responsible for
the two-mode squeezing 
\cite{Artoni1991,Artoni1989,Schwendimann1992,Schwendimann1992a,quattropani05,Ciuti2005PRB}.

\begin{figure}[tbp]
\includegraphics[width=\linewidth]{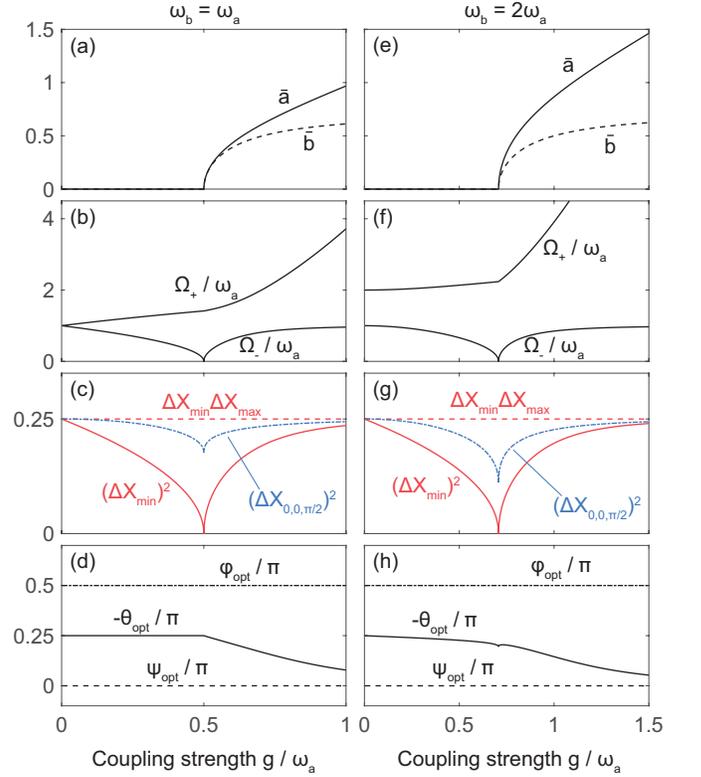}
\caption{For (a-d) $\omega_b = \omega_a$ and (e-h) $\omega_b = 2\omega_a$,
we plot, as a function of $g/\omega_a$, 
(a,e) order parameters $\ba$ and $\bb$,
(b,f) eigenfrequencies $\varOmega_{\pm}$,
(c,g) quadrature variance,
and (d,h) optimal angles $\thetaopt$, $\psiopt$, and $\phiopt$
that give the minimum variance $(\dXmin)^2$
[red solid curves in (c,g)].
The minimum variance vanishes
at the SRPT critical point
($\rabi = \sqrt{\omega_a\omega_b}/2$)
under satisfying the equality in the Heisenberg uncertainty principle
$\dXmin\dXmax = 1/4$ [red dashed line in (c,g)]
with the variance $(\dXmax)^2$ conjugate to $(\dXmin)^2$.}
\label{fig:1}
\end{figure}

Since the Dicke model can be treated effectively as an infinite-dimensional system
\cite{Larson2017JPA} in the thermodynamic limit ($N\to\infty$),
the SRPT can be analyzed under the mean-field framework
\cite{Wang1973PRA,Emary2003PRE,Emary2003PRL,Sharma2020}.
In the present letter, we follow the Holstein--Primakoff-transformation approach
\cite{Emary2003PRE,Emary2003PRL,Sharma2020},
which is suited for zero-temperature analyses.
The spin operators are rewritten by
a bosonic annihilation operator $\ob$ of the atomic collective excitations as
\begin{align}
\oS_z & \to \obd\ob- N/2, &
\oS_- & \to (N-\obd\ob)^{1/2} \ob.
\end{align}
The appearance of the superradiant phase,
where non-zero $\braket{\oa} = \sqrt{N}\ba$ and $\braket{\ob}=-\sqrt{N}\bb$
($\ba,\bb\in\mathbb{R}$) appear spontaneously, at the zero temperature
can be easily confirmed through the classical energy
$\bar{\HH}/(\hbar N) = \omega_a\ba^2 + \omega_b\bb^2 - 4\rabi\ba\bb\sqrt{1-\bb^2}$
obtained from Eq.~\eqref{eq:oHH_Dicke}.
The zero-temperature classical state satisfies
$\partial\HH/\partial\ba = \partial\HH/\partial\bb = 0$,
from which we have
\begin{align}
\ba & = \frac{2\rabi}{\omega_a}\bb\sqrt{1-\bb^2},&
\bb^2
& = \begin{cases}
      0, & \rabi \le \sqrt{\omega_a\omega_b}/2 \\
      \frac{1}{2}\left( 1 - \frac{\omega_a\omega_b}{4\rabi^2} \right), & \rabi > \sqrt{\omega_a\omega_b}/2
    \end{cases}
\end{align}
These are plotted as a function of $\rabi/\omega_a$
in Fig.~\ref{fig:1}(a) and (e) with
$\omega_b = \omega_a$ and $\omega_b = 2\omega_a$, respectively.
The zero-temperature SRPT occurs at 
\begin{equation} \label{eq:critical_point} 
\rabi = \sqrt{\omega_a\omega_b}/2,
\end{equation}
i.e., in the ultrastrong coupling regime \cite{Ciuti2005PRB,Kockum2018,Forn-Diaz2018}.

The quantum fluctuations around the zero-temperature classical state are described
by replacing $\oa$ and $\ob$ with $\sqrt{N}\ba+\oa$ and $-\sqrt{N}\bb+\ob$,
respectively \cite{Emary2003PRE,Emary2003PRL,Sharma2020}. After this replacement, $\hat{a}$ and $\hat{b}$ are now considered as the fluctuation operators.
The Dicke Hamiltonian, Eq.~\eqref{eq:oHH_Dicke}, is then expanded as
\begin{align} \label{eq:Hopfield} 
\oHH/\hbar & \equiv \omega_a\oad\oa + \wbt\obd\ob + \rabit(\oad+\oa)(\obd+\ob)
+ \DDt(\obd+\ob)^2
\nonumber \\ & \quad + O(N^{-1/2}) + \const,
\end{align}
where the coefficients are modified by the order parameters
$\ba$ and $\bb$ as
\begin{align}
\rabit & \equiv \frac{\rabi(1-2\bb^2)}{\sqrt{1-\bb^2}},&
\DDt & \equiv \frac{\rabi\ba\bb}{\sqrt{1-\bb^2}},&
\wbt & \equiv \omega_b + 2\DDt.
\end{align}
In the following, we consider the thermodynamic limit ($N\to\infty$)
and focus only on the leading terms in Eq.~\eqref{eq:Hopfield}, which gives rise to a quadratic Hamiltonian in terms of $\hat{a}$ and $\hat{b}$.

By describing the photonic and atomic fluctuations using Eq.~\eqref{eq:Hopfield},
we first demonstrate the perfect intrinsic two-mode squeezing numerically.
We consider a general superposition of the two fluctuation operators
defined with two angles $\theta$ and $\psi$ as
\begin{equation} \label{eq:oc} 
\oc_{\theta,\psi} \equiv \oa \cos\theta + \ee^{\ii\psi} \ob \sin\theta.
\end{equation}
We define a quadrature \cite{meystre99,walls08}
by this bosonic operator with a phase $\varphi$ as
\begin{equation} \label{eq:d_C1} 
\oX_{\theta,\psi,\varphi} = ({\oc_{\theta,\psi}\ee^{\ii\varphi} + \ocd_{\theta,\psi}\ee^{-\ii\varphi}})/{2}.
\end{equation}
We evaluate the variance
$(\dX_{\theta,\psi,\varphi})^2
\equiv \brag (\oX_{\theta,\psi,\varphi})^2 \ketg - \brag \oX_{\theta,\psi,\varphi} \ketg^2
= \brag (\oX_{\theta,\psi,\varphi})^2 \ketg$
of this quadrature
with respect to the ground state $\ketg$ of the fluctuation Hamiltonian, Eq.~\eqref{eq:Hopfield}.
Here, we consider annihilation operators $\op_{\pm}$ of eigenmodes
(i.e., polariton modes) that diagonalize Eq.~\eqref{eq:Hopfield} as
\begin{equation} \label{eq:oHH_ani=Omega_pd_p}
\oHH/\hbar = \varOmega_-\opd_-\op_- + \varOmega_+\opd_+\op_+ + O(N^{-1/2})+ \const,
\end{equation}
where $\varOmega_{\pm}$ are the eigenfrequencies.
The ground state $\ketg$ is determined by requiring
\begin{equation} \label{eq:p|0>=0} 
\op_{\pm}\ket{0} = 0.
\end{equation}
Due to the presence of the counter-rotating terms
$\oa\ob$, $\oad\obd$, $\ob\ob$, and $\obd\obd$,
originating from those in the Dicke model in Eq.~\eqref{eq:oHH_Dicke},
the eigenmode operators are obtained by a Bogoliubov transformation
\cite{Artoni1991,Artoni1989,Schwendimann1992,Schwendimann1992a,quattropani05,Ciuti2005PRB,Sharma2020} as
\begin{equation} \label{eq:op=wxyz_ob_oa}
\op_{\pm} = w_{\pm} \oa + x_{\pm} \ob + y_{\pm} \oad + z_{\pm} \obd.
\end{equation}
For positive eigenfrequencies $\varOmega_{\pm} > 0$, the coefficients must satisfy
$|w_{\pm}|^2 + |x_{\pm}|^2 - |y_{\pm}|^2 - |z_{\pm}|^2 = 1$
in order to yield $[\op_{\pm}, \opd_{\pm}] = 1$.
These coefficients and $\varOmega_{\pm}$ are determined
by an eigenvalue problem \cite{Ciuti2005PRB}
derived from Eq.~\eqref{eq:Hopfield} as
\begin{equation} \label{eq:eigenvalue_problem} 
\begin{pmatrix}
\omega_a & \rabit & 0 & -\rabit \\
\rabit & \wbt+2\DDt & -\rabit & -2\DDt \\
0 & \rabit & -\omega_a & -\rabit \\
\rabit & 2\DDt & -\rabit & -\wbt-2\DDt
\end{pmatrix}
\begin{pmatrix} w_{\pm} \\ x_{\pm} \\ y_{\pm} \\ z_{\pm} \end{pmatrix}
= \varOmega_{\pm}
\begin{pmatrix} w_{\pm} \\ x_{\pm} \\ y_{\pm} \\ z_{\pm} \end{pmatrix}.
\end{equation}
Two positive eigenvalues correspond to the eigenfrequencies $\varOmega_{\pm}$.
We also get two negative eigenvalues $-\varOmega_{\pm}$,
whose eigenvectors correspond to the creation operators $\opd_{\pm}$.
In this letter, we suppose
$0 \le \varOmega_- \le \varOmega_+$
i.e., $\varOmega_-$ and $\varOmega_+$ are the eigenfrequencies
of the lower and upper eigenmodes, respectively.
Figs.~\ref{fig:1}(b,f) show $\varOmega_{\pm}$ as functions of $\rabi/\omega_a$.
It is known \cite{Emary2003PRL,Emary2003PRE}
that the lower eigenfrequency $\varOmega_{-}$ vanishes
at the SRPT critical point $\rabi = \sqrt{\omega_a\omega_b}/2$.
In this case, $[\op_-, \opd_-] = 1$ does not hold, because Eq.~\eqref{eq:eigenvalue_problem} gives two degenerated solutions with $\varOmega_- = 0$ mathematically. In the following, we will show that perfect squeezing is obtained at this critical point.

The quadrature variance $(\dX_{\theta,\psi,\varphi})^2 = \brag (\oX_{\theta,\psi,\varphi})^2 \ketg$
can be evaluated by rewriting the original photonic and atomic fluctuation operators
$\oa$, $\oad$, $\ob$, and $\obd$ with the eigenmode operators $\op_{\pm}$
and $\opd_{\pm}$ and using Eq.~\eqref{eq:p|0>=0}.
We numerically searched for the optimal angles $\thetaopt$, $\psiopt$, and $\phiopt$
that give the minimum variance
$(\dXmin)^2 \equiv (\dX_{\thetaopt,\psiopt,\phiopt})^2$
for given $\omega_a$, $\omega_b$, and $\rabi$.

In Fig.~\ref{fig:1},
(c,g) quadrature variances including $(\dXmin)^2$
and (d,h) optimal angles $\thetaopt$, $\psiopt$, and $\phiopt$
are plotted as functions of $\rabi/\omega_a$
for (c,d) $\omega_b = \omega_a$ and (g,h) $\omega_b = 2\omega_a$.
As shown by red solid lines in Figs.~\ref{fig:1}(c,g),
while the minimum variance is $(\dXmin)^2 = 1/4$
(standard quantum limit \cite{meystre99,walls08})
in the absence of the photon--atom coupling ($\rabi = 0$),
it decreases as $\rabi$ increases
and vanishes at the SRPT critical point
$\rabi = \sqrt{\omega_a\omega_b}/2$.
After that, in the superradiant phase ($\rabi>\sqrt{\omega_a\omega_b}/2$),
$(\dXmin)^2$ increases again and approaches $1/4$ asymptotically.
The variance of its conjugate fluctuation is given by
$(\dXmax)^2 \equiv (\dX_{\thetaopt,\psiopt,\phiopt-\pi/2})^2$,
which diverges at the SRPT critical point
(not shown in the figure).
However, as shown by red dashed lines in Figs.~\ref{fig:1}(c,g),
we numerically confirmed that the product
of these variances satisfy $\dXmin\dXmax = 1/4$,
the equality in the Heisenberg uncertainty principle,
i.e., an ideal two-mode squeezing is obtained.

In Figs.~\ref{fig:1}(c,g),
the blue dash-dotted curves represent the variance
$(\dX_{0,0,\pi/2})^2 = \brag (\oa-\oad)^2 \ketg/4$ of a photonic fluctuation.
Such a one-mode variance does not vanish even at the critical point
\cite{Emary2003PRE,Castanos2011,Shapiro2019}
and satisfies only the inequality
$\dX_{0,0,\pi/2}\dX_{0,0,0} > 1/4$ in the Heisenberg uncertainty principle
(not shown in the figure).

As seen in Figs.~\ref{fig:1}(d,h),
in the present case,
the minimum variance is obtained always for
$\psiopt = 0$ (dashed line) and $\phiopt = \pi/2$ (dash-dotted line).
These two phases depend on those of the coupling strengths
of the co-rotating and counter-rotating terms \cite{Note4},
although we simply considered
the isotropic Dicke model, Eq.~\eqref{eq:oHH_Dicke},
and real $\rabi$ in the present letter.
On the other hand, $\thetaopt$ (solid curves)
depend on $\rabi/\omega_a$ and $\omega_b/\omega_a$ in general,
while $\thetaopt = -\pi/4$,
i.e., $(\dX_{-\pi/4,0,\pi/2})^2 = - \brag (\oa-\ob-\oad+\ob)^2 \ketg/8$
always gives the minimum variance
in the normal phase ($\rabi < \omega_a/2)$
for $\omega_b = \omega_a$.


Next, we try to understand the numerically found perfect and ideal squeezing
($\dXmin = 0$ at the critical point with $\dXmin\dXmax = 1/4$)
by an analytical expression
of the ground state $\ketg$ of the fluctuation Hamiltonian, Eq.~\eqref{eq:Hopfield}.
Following the discussion by Schwendimann and Quattropani
\cite{Schwendimann1992,Schwendimann1992a,quattropani05},
we consider a unitary operator $\oU$
that transforms the fluctuation operators $\oa$ and $\ob$
into the eigenmode ones $\op_{\pm}$ as
\begin{subequations} \label{eq:op_pm} 
\begin{align}
\op_- & \equiv \oU \oa \oUd, &
\op_+ & \equiv \oU \ob \oUd.
\end{align}
\end{subequations}
For the vacuum $\ket{0_{a,b}}$ of the individual fluctuations
satisfying $\oa\ket{0_{a,b}} = \ob\ket{0_{a,b}} = 0$,
the ground state $\ketg$ of the coupled system can be expressed as
\begin{equation} \label{eq:|0>} 
\ket{0} \propto \oU \ket{0_{a,b}},
\end{equation}
while there is a freedom of introducing an overall phase factor.
This expression certainly satisfies Eq.~\eqref{eq:p|0>=0}.

The explicit expression of $\oU$ for the fluctuation Hamiltonian, Eq.~\eqref{eq:Hopfield},
derived from the Dicke model
has been shown recently by Sharma and Kumar \cite{Sharma2020} as
\begin{equation} \label{eq:oU_Sharma} 
\oU \equiv \oUz \oU_- \oU_+,
\end{equation}
where the three unitary operators are defined as
\begin{subequations} \label{eq:oU_b12+-} 
\begin{align}
\oUz & \equiv \ee^{-(r_b/2)(\obd\obd-\ob\ob)} \ee^{-\alpha(\oad\ob-\obd\oa)} \ee^{-r(\oad\obd-\ob\oa)}, \label{eq:oU_b} \\
\oU_- & \equiv \ee^{-(r_-/2)(\oad\oad-\oa\oa)},\ 
\oU_+ \equiv \ee^{-(r_+/2)(\obd\obd-\ob\ob)}.
\end{align}
\end{subequations}
$\oU_{\pm}$ are one-mode squeezing operators,
and $\oUz$ is a product of one-mode squeezing, superposing, and two-mode squeezing operators
\cite{meystre99,walls08}.
By a Bogoliubov transformation of $\ob$ for renormalizing the $\DDt$ term
in Eq.~\eqref{eq:Hopfield},
the atomic frequency and coupling strength are modified again as
\begin{align}
\wbc & \equiv \sqrt{\wbt(\wbt+4\DDt)}, &
\rabic & \equiv \sqrt{({1-\gamma})/({1+\gamma})}\rabit,
\end{align}
where $\gamma$, giving also $r_b$ in Eq.~\eqref{eq:oU_b}, is defined as
\begin{equation}
\gamma \equiv \frac{\sqrt{1+4\DDt/\wbt}-1}{\sqrt{1+4\DDt/\wbt}+1}
= \tanh(r_b).
\end{equation}
The other factors in Eqs.~\eqref{eq:oU_b12+-} are defined as
\begin{subequations} \label{eq:alpha_r_r_pm} 
\begin{align}
\tan(2\alpha) & = {2\rabic}/({\omega_a-\wbc}), \label{eq:tan2alpha} \\
\tanh(2r) & = {2\rabic\cos(2\alpha)}/({\omega_a+\wbc}), \\
\tanh(2r_-) & = {\rabic\sin(2\alpha)}/{\epsilon_-}, \label{eq:tanh(2r-)} \\
\tanh(2r_+) & = - {\rabic\sin(2\alpha)}/{\epsilon_+},
\end{align}
\end{subequations}
where $\epsilon_{\pm}$ and the eigenfrequencies $\varOmega_{\pm}$ are expressed as
\begin{align}
\epsilon_{\pm}
& \equiv \sqrt{ \frac{(\omega_a+\wbc)^2}{4} - \rabic^2\cos^2(2\alpha) }
  \pm \sqrt{ \frac{(\omega_a-\wbc)^2}{4} + \rabic^2 }, \label{eq:epsilon} \\ 
\varOmega_{\pm}
& = \sqrt{\epsilon_{\pm}{}^2-\rabic^2\sin^2(2\alpha)}.
\label{eq:Omega} 
\end{align}
Note that the unitary operator $\oU$ can be rewritten as
\begin{equation} \label{eq:oU_e} 
\oU = \oU_{d-} \oU_{d+} \oUz,
\end{equation}
i.e., a product of $\oUz$ and two one-mode squeezing operators
\begin{equation}
\oU_{d\pm}
\equiv \oUz \oU_{\pm} \oUzd
= \ee^{(-r_{\pm}/2)(\odd_{\pm}\odd_{\pm}-\od_{\pm}\od_{\pm})}
\end{equation}
under a new basis transformed from the original one ($\oa$ and $\ob$) by $\oUz$ as
\begin{align}
\od_- & \equiv \oUz \oa \oUzd, &
\od_+ & \equiv \oUz \ob \oUzd. \label{eq:od+-} 
\end{align}

In the case of $\omega_a = \omega_b$ and in the normal phase
($\rabi<\sqrt{\omega_a\omega_b}/2$, $\ba = \bb = r_b = \gamma = 0$, $\wbc = \omega_b$,
and $\rabic = \rabi$),
we can easily find that the ground state
$\ketg \propto \oU\ket{0_{a,b}}$ is an ideal two-mode squeezed vacuum.
From Eqs.~\eqref{eq:alpha_r_r_pm}, \eqref{eq:epsilon}, and \eqref{eq:Omega},
in the limit of $\omega_b \to \omega_a + 0^+$,
we get $\varOmega_{\pm} = \sqrt{\omega_a(\omega_a\pm2\rabi)}$,
$\alpha = - {\pi}/{4}$,
$r = 0$,
$\tanh(2r_-) = - {\rabi}/({\omega_a-\rabi})$,
and $\tanh(2r_+) = {\rabi}/({\omega_a+\rabi})$.
Since the unitary operator $\oUz$ is simply
a superposing operator as $\oUz = \ee^{(\pi/4)(\oad\ob-\obd\oa)}$,
the new basis $\od_{\pm}$ defined in Eq.~\eqref{eq:od+-}
are just the equal-weight superpositions of the original fluctuation operators as
$\od_{\pm} = ({\oa\pm\ob})/{\sqrt{2}}$.
Then, the ground state is simply expressed as
$\ketg \propto \oU\ket{0_{a,b}} = \oU_{d-} \oU_{d+} \ket{0_{a,b}}$,
i.e., squeezed by $r_{\pm}$ in the two-mode (superposed) basis $\od_{\pm}$,
and the variances of quadratures defined 
by $\od_{-} = \oc_{-\pi/4,0}$ are obtained as
$(\dXmin)^2 = (\dX_{-\pi/4,0,\pi/2})^2 = \ee^{2r_-}/4$
and $(\dXmax)^2 = (\dX_{-\pi/4,0,0})^2 = \ee^{-2r_-}/4$.
Then, $\dXmin\dXmax=1/4$ is satisfied for any $\rabi$.
When the coupling strength reaches the critical point as
$\rabi \to \omega_a/2 + 0^-$,
the lower eigenfrequency becomes $\varOmega_- \to 0^+$,
and the perfect squeezing is obtained as
$r_- \to - \infty$ in the $\od_-$ basis.
Therefore, the quadrature variance
$(\dXmin)^2$ vanishes
at the SRPT critical point,
as we demonstrated in Fig.~\ref{fig:1}.

In the general case with $\omega_a \neq \omega_b$ case
(and in the superradiant phase),
we can mathematically confirm that perfect squeezing can be obtained
from the expression $\ketg \propto \oU\ket{0_{a,b}}$ of the ground state  described
by the unitary operator $\oU$ in Eq.~\eqref{eq:oU_e},
while the basis $\od_{\pm}$ is not simple superpositions of the original fluctuation
operators $\oa$ and $\ob$ but includes also their creation operators $\oad$ and $\obd$.
Instead of such a straightforward but complicated analysis,
we can understand the perfect squeezing
at the SRPT critical point $\rabi=\sqrt{\omega_a\omega_b}/2$
in the following manner.

The perfect squeezing can be obtained generally when the quadrature
$\oX_{\theta,\psi,\varphi}
= [(\ee^{\ii\varphi}\oa +\ee^{-\ii\varphi}\oad) \cos\theta
+ \ee^{\ii\psi}(\ee^{\ii\varphi} \ob +\ee^{-\ii\varphi}\obd) \sin\theta])/{2}$
is proportional to the eigenmode operator $\op_-$,
because $\op_-\ketg = 0$ and then
the quadrature variance $\braket{0|(\oX_{\theta,\psi,\varphi})^2|0}$ becomes zero.
Since we can freely choose the angles $\theta$, $\psi$, and $\varphi$,
the perfect squeezing can be obtained
when the weights of the annihilation and creation operators
in the eigenmode operator $\op_- = w_-\oa + x_-\ob + y_-\oad + z_-\obd$
are equal as $|w_-| = |y_-|$ and $|x_-| = |z_-|$.
Such equal weights are obtained at critical points
accompanied by a vanishing resonance frequency
in some interacting systems,
e.g., weakly interacting Bose gases \cite{Fetter1972}.
In the present case, we can easily find that
$w_-/y_-= x_-/z_- = -1$ is obtained for $\varOmega_- = 0$
from the eigenvalue problem in Eq.~\eqref{eq:eigenvalue_problem}.
In this way, we can generally get perfect squeezing in a proper quadrature
at critical points in the Dicke model
and also in similar models with counter-rotating terms
and a vanishing resonance frequency.


In summary, we found that perfect and ideal squeezing
is an intrinsic property associated with the zero-temperature SRPT in the Dicke model.
Phenomenologically,
owing to a possible divergence of quantum fluctuation at a critical point,
its conjugate fluctuation can be perfectly squeezed
under satisfying the Heisenberg uncertainty principle.
Such an ideal quantum behavior should be obtained only in limited systems
with a vanishing resonance frequency and counter-rotating terms,
and we confirmed that the Dicke model is one of such systems.

In contrast to the standard squeezing generation processes
in dynamical and nonequilibrium situations \cite{meystre99,walls08},
the phenomenon of intrinsic squeezing we described here does not diminish in time and is stably obtained in equilibrium situations.
While perfect intrinsic spin squeezing has been reported in some spin models
such as the Lipkin--Meshkov--Glick model \cite{Ma2009}, the XY model \cite{Liu2013},
and the transverse-field Ising model \cite{Frerot2018},
this work presented the first photon--matter coupled model in which perfect intrinsic squeezing arises.
Intrinsic squeezing has a potential for improving continuous-variable quantum information technologies
\cite{Braunstein2005,Adesso2014},
which have been developed mostly in photonic systems,
by making them more resilient to decoherence.
For practical applications, including quantum metrology \cite{Frerot2018},
intrinsic squeezing at finite temperatures,
for finite number ($N$) of atoms,
and in the presence of coupling with a bath
should be investigated in the future.

\begin{acknowledgments}
M.B.~acknowledges support from the JST PRESTO program (grant JPMJPR1767).
J.K.~acknowledges support from
the U.S.\ National Science Foundation (Cooperative Agreement DMR-1720595),
and the U.S.\ Army Research Office (grant W911NF-17-1-0259).
H.P.~acknowledges support from the U.S. NSF and the Welch Foundation (Grant No. C-1669).
We thank Tomohiro Shitara for a fruitful discussion.
\end{acknowledgments}


\end{document}